# Absence of superconductivity in ultra-thin layers of FeSe synthesized on a topological insulator

Andreas Eich[1], Nils Rollfing[1], Fabian Arnold[2], Charlotte Sanders[2], Pascal R. Ewen[1], Marco Bianchi[2], Maciej Dendzik[2], Matteo Michiardi[2], Jian-Li Mi[3], Martin Bremholm[4], Daniel Wegner[1], Philip Hofmann[2] and Alexander A. Khajetoorians[1*]

[1] Institute for Molecules and Materials, Radboud University, 6525 AJ Nijmegen, Netherlands

[2] Department of Physics and Astronomy, Interdisciplinary Nanoscience Center, Aarhus University, 8000 Aarhus C, Denmark

[3] Institute for Advanced Materials, School of Materials Science and Engineering, Jiangsu University, Zhenjiang 212013, Jiangsu, P.R. China

[4] Center for Materials Crystallography, Department of Chemistry and iNANO, Aarhus University, 8000 Aarhus C, Denmark

[*] Correspondence to: a.khajetoorians@science.ru.nl

The structural and electronic properties of FeSe ultra-thin layers on $Bi_2Se_3$ have been investigated with a combination of scanning tunneling microscopy and spectroscopy and angle-resolved photoemission spectroscopy. The FeSe multi-layers, which are predominantly 3-5 monolayers (ML) thick, exhibit a hole pocket-like electron band at $\bar{\Gamma}$ and a dumbbell-like feature at $\bar{M}$, similar to multi-layers of FeSe on $SrTiO_3$. Moreover, the topological state of the $Bi_2Se_3$ is preserved beneath the FeSe layer, as indicated by a heavily *n*-doped Dirac cone. Low temperature STS does not exhibit a superconducting gap for any investigated thickness down to a temperature of 5 K.

**Introduction**

Iron-based superconductors show some types of unusual high $T_C$ behavior, as well as a competition between spin-density wave (SDW) and antiferromagnetic order [1, 2]. The recent remarkable finding that a single layer of FeSe grown on $SrTiO_3$ exhibits $T_C$ values above 100 K, compared to the bulk value of 8 K, has created a strong interest in unconventional superconductivity in ultra-thin films [3-7]. An extensive number of studies has probed the role of the interface, film thickness, and doping on the phase of FeSe thin films [8-11], illustrating that the electronic properties and consequent superconductivity can be strongly modified by tailoring the properties of the FeSe films. While it is not clear what makes $SrTiO_3$ the interface of choice, high temperature superconductivity in a single layer of FeSe seems to be accompanied by the presence of electron pockets at the $\bar{M}$-point and the disappearance of a hole pocket at $\bar{\Gamma}$ of the Fermi surface [8]. If and how the pairing mechanism is modified as a function of thickness, especially in the regime of a few monolayers (ML), and where the transition from thin film to bulk-like superconductivity occurs, are questions of major interest [8, 12].

In addition to high temperature superconductivity, superconductivity in ultra-thin films is a route toward topological superconductivity [6, 13] for example, via vortex states formed near the interface of a topological insulator [14]. However, it is relatively difficult to grow epitaxial layers on prototypical topological insulators such as the bismuth chalcogenides [15], and this makes it challenging to interface superconductors with well-known topological insulators. Recently, it was shown that FeSe can be synthesized directly on $Bi_2Se_3$ by heating $Bi_2Se_3$ crystals after depositing overlayers of Fe [16]. Such an interface may be an intriguing system for investigations of topological superconductivity. However, to date no study has investigated the superconductivity of these FeSe layers on $Bi_2Se_3$.

We present a detailed study of the structural and electronic properties of FeSe ultra-thin films synthesized on $Bi_2Se_3$. We use scanning tunneling microscopy/spectroscopy (STM/STS) to characterize the electronic properties of these FeSe layers, and compare these results to those obtained by angle-resolved photoemission spectroscopy (ARPES). We see no evidence of a superconducting gap at a temperature of 5 K. The topological state remains intact at the interface between the FeSe and the $Bi_2Se_3$, but shifts down in energy 200 meV relative to pristine $Bi_2Se_3$. We corroborate the quenching of superconductivity with the absence of the characteristic ring-like electron pocket at the $\bar{M}$-point that is seen for thin superconducting layers of FeSe on $SrTiO_3$.

**Experimental details**

All experiments were carried out in ultra-high vacuum conditions. $Bi_2Se_3$ crystals were synthesized in the manner that has been described previously [17, 18]. The crystals were cleaved *in situ* with scotch tape to create a clean surface. Fe was deposited at room temperature from an electron beam evaporator, and the sample was subsequently annealed at 570 K for 45 min in order to form FeSe films, in a manner similar to that of Ref. [16]. We define the coverage here as the percent (%) areal coverage of FeSe on $Bi_2Se_3$, as characterized with STM.

STM/STS were conducted with an Omicron LT-STM operated at 5 K, with the bias applied to the sample. An electrochemically etched tungsten wire, flashed *in situ,* was used as the tip. Spectroscopy was performed utilizing a lock-in technique and recording the differential conductance, with modulation frequencies between $f_{mod}$ = 600 and 870 Hz. An overall effective electron temperature of 6 K was confirmed by probing the superconducting gap of Pb(111). Topography images were taken in constant-current mode. ARPES data were acquired at the SGM3 beamline of the ASTRID2 synchrotron light source at a temperature of 90 K and with beam energies between $h\nu$ = 18 eV and 53.6 eV [19]. LEED images were taken at a temperature of 90 K and at an electron kinetic energy of $E_{kin}$ = 50.2 eV.

**Structural characterization**

Fig. 1(a) shows a typical topography image with a coverage of 65% FeSe on $Bi_2Se_3$. As we discuss below and as previously reported [16], excess Se during heating of the $Bi_2Se_3$ bonds with Fe deposited on the surface to form ultra-thin layers of FeSe; the FeSe/$Bi_2Se_3$ interface is typically below the level of the surrounding bare $Bi_2Se_3$ (Fig. 1). The signature of FeSe is a linear stripe moiré pattern with a periodicity of 6.7±0.7 nm. The moiré pattern originates from lattice mismatch between the FeSe lattice and underlying $Bi_2Se_3$ lattice. The moiré structure can be found in three different orientations rotated by angles of 120°. The height difference between terraces exhibiting this pattern is $c_{FeSe}$ = 0.57±0.02 nm (Fig. 1(b)). Moreover, the intensity of this moiré pattern decreases as the thickness of the FeSe layer is increased (see, higher islands in Fig. 1(a)). Atomic resolution images of the FeSe areas (Fig. 1(d)) reveal a square unit cell with a lattice constant of $a_{FeSe}$ = 0.38±0.01 nm, in contrast to the areas we identify as $Bi_2Se_3$ (Fig. 1(c)) which exhibit the expected three-fold symmetric unit cell ($a_{Bi2Se3}$ = 0.418±0.06 nm). The terrace height between different thicknesses of FeSe and the measured lateral lattice constant are close to the bulk values of FeSe [5, 7], and identical to previously investigated layers of FeSe on $Bi_2Se_3$ [16].

The apparent height difference between FeSe islands of the minimal observed height and $Bi_2Se_3$ is roughly 0.3 nm. It is important to note that this measured difference does not vary strongly with the applied bias voltage (Fig 1(b)). From the measured apparent heights, we can rule out the possibility that we have one single ML of FeSe on top of $Bi_2Se_3$; rather, the FeSe layers interface with the $Bi_2Se_3$ at a level beneath the adjacent exposed $Bi_2Se_3$ surface, as also concluded in ref. [16]. Although we cannot precisely determine the number of FeSe layers which exist beneath the surface based on our STM results alone, we can nevertheless take into consideration (i) the amount of deposited Fe and the measured surface coverage, and (ii) the appearance of the moiré structure; from these we can conclude that we are in the ultra-thin limit with, presumably, approximately 3-5 ML of FeSe [16]. Here we focus on the thinnest layer (characterized also by the greatest intensity variation in the moiré pattern), which is also the predominant thickness observed in the present study.

Two types of defects are found on the FeSe films, referred to as types A and B. Type A defects appear as two bright spots on neighboring atom sites, and have been previously identified as a Se atom on or close to an Fe site [20-22] or an Fe vacancy [23]. Type B defects appear as depressions at an atomic site at the top of the film. The structure of type B defects is unknown.

The $Bi_2Se_3$ regions exhibit large-scale quasi-hexagonal regions between 20-40 nm in length, separated by darker regions characterized by variation in apparent height between the quasi-hexagonal regions (Fig. 1(a)); a similar observation has been made in a previous study [16]. We note that similar structures were previously also observed in binary alloys stemming from a buried dislocation network [24]. As Se migration during heating is responsible for forming the FeSe layers, we associate this structure with a dislocation network that penetrates into the $Bi_2Se_3$ surface.

The locally measured atomic and moiré lattices of the FeSe are corroborated by LEED measurements: see Fig. 1(e-f) for LEED images taken on the bare $Bi_2Se_3$ (e) and after the preparation of FeSe (f). The LEED data after preparation exhibits a square pattern for each of the three 120°-rotated domains (indicated by green, orange and purple arrows). The linear moiré leads to satellite spots at each corner of the squares. The lattice constants calculated from the LEED pattern are $a_{FeSe}$ = 0.38±0.01 nm for the atomic lattice and 6.3±0.2 nm for the moiré pattern, in agreement with the STM measurements.

**Electronic structure characterization**

ARPES spectra are shown in Figs. 2 and 3. Photon energies of 18.0, 26.0 and 53.6 eV were chosen so special features of FeSe or $Bi_2Se_3$ could be enhanced, exploiting the energy-dependent matrix elements in the photoemission processes. The light beam spot has a diameter on the order of 100 µm [19], so it simultaneously probes regions of both bare substrate and islands; thus, the photoemission intensity contains a superposition of signals from both. Fig. 2(a) shows the photoemission intensity along the high-symmetry direction $\bar{\Gamma}$-$\bar{M}_{Bi2Se3}$ (where $\bar{\Gamma}$ denotes the center of the primary surface Brillouin zone (BZ), and the indices of the high-symmetry point indicates the material). There are two Dirac cones visible, corresponding to two co-existing domains of $Bi_2Se_3$ with different degrees of doping. Comparison with data acquired from bare $Bi_2Se_3$ [25] (not shown) suggests that the less *n*-doped state is that of the pristine $Bi_2Se_3$ surface and rules out the structural scenario that a Bi bilayer is situated on top of $Bi_2Se_3$ [15]. We assign the more *n*-doped state to the persistent $Bi_2Se_3$ surface state at the buried interface beneath FeSe islands. Enhanced *n*-doping of the substrate at the buried interface is presumably accompanied by electron depletion in the FeSe overlayer. Given the very short inelastic mean free path for photoelectrons here, the observation of the Dirac cone at the interface confirms the fact that the FeSe layers must be very thin. Both Dirac cones exhibit a hexagonal warping, as can be seen in a map of the photoemission intensity at the Fermi energy (Fig. 2(b)). In addition to the two Dirac cones, Fig. 2(a) exhibits a broad, relatively flat band, which is marked with a blue arrow, at a binding energy of $E_{bin} \approx 0.2$ eV; the same feature is marked with an identical blue arrow in Fig. 3(b) and (c) (where, however, it appears more intense, due to the different photon energy used to acquire data in that case). It seems to be associated with the flat, broad band that stretches along the high symmetry directions of the FeSe Brillouin zone ($\bar{\Gamma}$-$\bar{X}_{FeSe}$-$\bar{M}_{FeSe}$-$\bar{\Gamma}$). It is not present in data acquired on bare $Bi_2Se_3$, and thus can be ascribed to FeSe. This band resembles a similar feature previously obtained in calculations of free-standing monolayer FeSe, where it was attributed to Fe 3*d* orbitals [11, 26].

Fig. 2(c) shows a constant energy contour, in which the spectral weight has been integrated from a binding energy +0.055 to -0.055 eV to enhance the photoemission intensity. In this range, a secondary $Bi_2Se_3$ BZ can be seen in addition to the primary BZ, and the corresponding $\bar{\Gamma}'_{Bi2Se3}$ can be identified by the dispersion. Knowing the positions of $\bar{\Gamma}$ and $\bar{\Gamma}'_{Bi2Se3}$, the hexagonal Brillouin zone of Bi$_2$Se$_3$ can be constructed. The orientations of the BZs of the three rotated FeSe domains relative to the primary Bi$_2$Se$_3$ BZ are known from the LEED measurements in Fig. 1(e) and (f), and this allows us to construct the FeSe BZs in Fig. 2(c). The primary and secondary Bi$_2$Se$_3$ BZs and the three primary FeSe BZs are indicated schematically with dashed lines in the figure. The three features in the second BZ are significantly weakened by matrix element effects, but are still visible. Because of the relatively small island size and the much larger size of the light beam spot, the combined photoemission intensity of all three FeSe domains is always simultaneously present in these measurements. Therefore, it is difficult to uniquely assign features of the photoemission spectrum to only one particular domain.

Fig. 3(a) shows the photoemission intensity at the Fermi surface, measured with a photon energy that enhances the appearance of FeSe features. Besides the signal from the bare and buried Bi$_2$Se$_3$ states at $\bar{\Gamma}$, another state with very low spectral weight is visible at $\bar{M}_{FeSe}$. The inset in Fig. 3(a) shows this feature with high contrast. The feature appears as a round intensity maximum to the left of $\bar{M}_{FeSe}$. By careful examination of the data shown in the inset, one can see a corresponding round feature in the neighboring Brillouin zone, such that both structures together form a dumbbell shape across the Brillouin zone boundary. The lobe in the second BZ is very weak due to the matrix element effects mentioned above. In order to reveal the origin of the dumbbell-like feature, the photoemission intensity along the $\bar{\Gamma}$- $\bar{X}_{FeSe}$-$\bar{M}_{FeSe}$-$\bar{\Gamma}$ directions is shown in Fig. 3(b). The faint feature near the Fermi level at $\bar{M}_{FeSe}$ that is responsible for the dumbbell-like feature is visible. An analysis method based on the mathematical concept of curvature can be used to locate maxima in photoemission intensity profiles and to emphasize weak features on an intense background. In order to see the faint feature near the Fermi level at $\bar{M}_{FeSe}$ more clearly, we thus plot the curvature of the data, as outlined in Ref. [27]. The resulting curvature plot in Fig. 3(c) identifies the dumbbell-like feature as a small electron pocket located slightly away from the $\bar{M}_{FeSe}$ point. Fig. 3(d-g) show the photoemission intensities and corresponding curvature plots for cuts through $\bar{\Gamma}$ and $\bar{M}_{FeSe}$, respectively, along the directions indicated in Fig. 3(a). At the photon energy used here, neither of the two Dirac cones is strongly visible at $\bar{\Gamma}$. Instead, a dome-shaped hole pocket is visible at $\bar{\Gamma}$. As this hole pocket is never observed for bare $Bi_2Se_3$, it is evidently associated with FeSe. Near $\bar{M}_{FeSe}$ the aforementioned small electron pocket, crossing the Fermi level slightly away from the high symmetry point, is visible.

Previous photoemission studies of FeSe films on $SrTiO_3$ shed light on these findings. Here, the superconducting state is associated with a circularly shaped feature around $\bar{M}_{FeSe}$ [28, 29]. By contrast, a dumbbell feature around $\bar{M}_{FeSe}$ and a dome-like hole pocket at $\bar{\Gamma}$ were observed for 3 ML of FeSe grown on $SrTiO_3$; the emergence of these features has been associated with a non-superconducting state, similar to what is seen here [8]. In agreement with our STM findings, our ARPES results further support the conclusion that we are probing ultra-thin FeSe layers thicker than 1 ML.

In order to probe the possible existence of superconductivity, we utilized high-resolution STS at $T = 5$ K (Fig. 4(a)). We do not observe a superconducting gap on any FeSe layers. We cross-checked the spectroscopic findings at various positions on a given terrace and on many terraces exhibiting different apparent layer heights for a given sample preparation. With more than 400 measurements on 11 different preparations and three different crystals, we did not observe a reproducible superconducting gap in STS.

For comparison with the ARPES characterization, we performed STS over a larger energy range. Fig. 4(b) shows large range spectra taken on the $Bi_2Se_3$ surface (red) and the FeSe layers (blue). The spectrum taken on the $Bi_2Se_3$ exhibits the typical features of that material [30]. Unlike the bare $Bi_2Se_3$, FeSe exhibits a characteristic broad peak centered around $V_S = -225$ mV, which corroborates the flatly dispersing band seen in Fig. 3. It is important to note that no differences are found for spectra taken on the two different FeSe thicknesses we observe.

While we do not observe a superconducting gap with STS, we do see the appearance of resonance states at low energy which are localized near defects of the FeSe films (Fig. 5). As we illustrate with a distance-dependent measurement, type A defects feature a resonance in the filled states centered around $V_S$= -6 mV. The resonance is clearly localized on the defect and not detectable when probed two atomic lengths away. Type B defects show a similar behavior, but with the resonance centered near $V_S$=19 mV (not shown).

**Conclusion**

In summary, we have grown ultra-thin layers of FeSe on $Bi_2Se_3$, and have characterized their structure and electronic properties. On the basis of structural characterization, we rule out the existence of single-layer FeSe films prepared by our method. Surprisingly, STS does not reveal a superconducting gap at $T = 5$ K. To this finding, ARPES measurements contribute additional insights into the reasons for this absence of superconductivity. A hole pocket at $\bar{\Gamma}$ and a dumbbell-like feature at $\bar{M}_{FeSe}$ at the Fermi surface are similar to the electronic structure of 3 ML undoped FeSe grown on $SrTiO_3$, which likewise does not exhibit superconductivity.[8] This is further evidence that we are probing a thickness greater than 1ML. We observe two Dirac cones, one being that of the exposed $Bi_2Se_3$, and one being a strongly *n*-doped Dirac cone which we attribute to the topological state at the FeSe/$Bi_2Se_3$ interface. Additionally, a flat band below the Fermi level is observed, similar to what has previously been predicted for freestanding single layers of FeSe [11, 26], and the existence of this band is corroborated by STS. The resemblance of the observed FeSe band structure to that seen for 3 ML FeSe on $SrTiO_3$ suggests that it might be possible here, as in that system, to push the FeSe/$Bi_2Se_3$ system into a superconducting state by *n*-doping the FeSe layers e.g., with K adatoms [8].

**Acknowledgments**

A.E., D.W., and A.A.K. gratefully acknowledge financial support from the Emmy Noether Program

(KH324/1-1) via the Deutsche Forschungsgemeinschaft, and from the Foundation for Fundamental Research on Matter (FOM), which is part of the Netherlands Organization for Scientific Research (NOW). A.A.K. also acknowledges the VIDI program (Project No. 680-47-534) from the Netherlands. P.H. acknowledges financial support from VILLUM FONDEN, the Danish Council for Independent Research, Natural Sciences under the Sapere Aude program (Grants No. DFF-4002-00029, and No. 0602-02566B), and the Lundbeck Foundation. M.B. acknowledges the financial support from Materials Crystallography (CMC), a Center of Excellence funded by the Danish National Research Foundation (DNRF93).

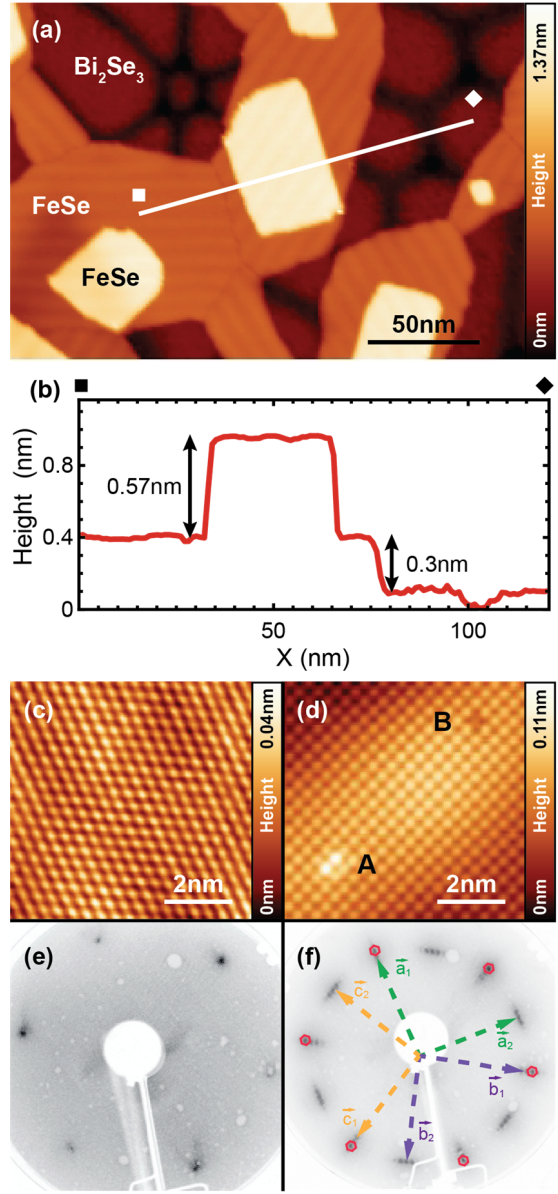

**Fig 1. (a)** Ultra-thin FeSe islands on $Bi_2Se_3$ exhibit a linear moiré pattern, whereas the $Bi_2Se_3$ surface exhibits a dislocation network. **(b)** Apparent height profile taken along marker in (a). The FeSe layers exhibit an apparent interlayer height of 0.57 nm; the apparent height difference between FeSe and $Bi_2Se_3$ is 0.3 nm. $V_S$ = 500 mV, $I_t$ = 300 pA. **(c)** The atomic lattice of $Bi_2Se_3$ observed in regions away from FeSe. $V_S$ = 500 mV, $I_t$ = 500 pA, $T$ = 5 K. **(d)** The FeSe patches exhibit a square lattice. Two different types of defects, A and B, can be identified. $V_S$ = 500mV, $I_t$ = 300pA. **(e)** LEED image

of the pristine Bi$_2$Se$_3$ showing the expected hexagonal pattern. **(f)** LEED after the deposition of Fe and annealing shows a superposition of several reciprocal lattices: in addition to the hexagonal pattern stemming from Bi$_2$Se$_3$, a square lattice appears in three domains each rotated by 120°. We assign these three domains to FeSe. The different domains are marked by arrows in green, orange and purple.

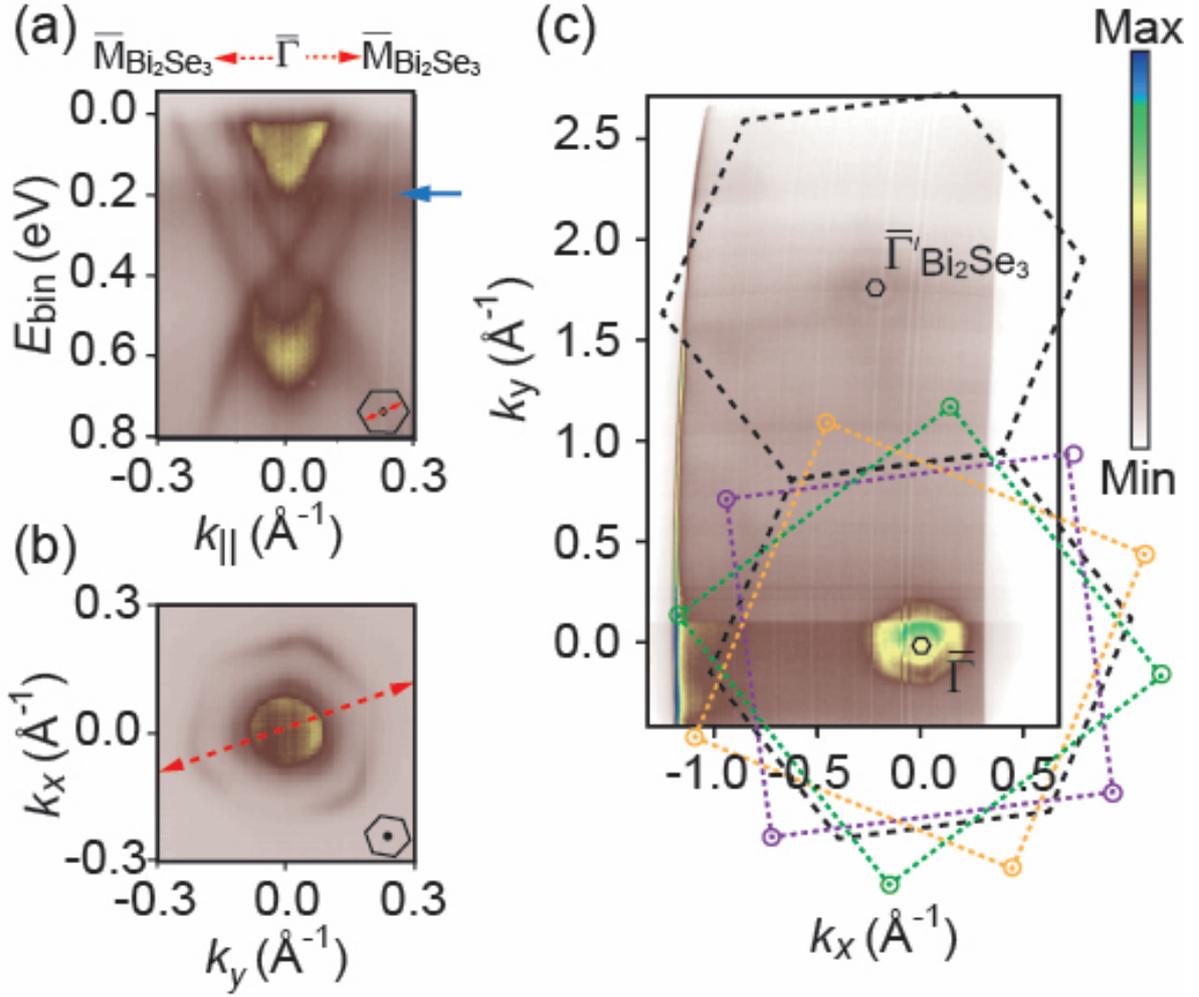

**Fig 2 (a)** Photoemission intensity map for FeSe/Bi$_2$Se$_3$. The red dashed arrow in the inset hexagon indicates the direction ($\bar{\Gamma}$-$\bar{M}_{Bi2Se3}$) of the cut through the surface BZ of Bi$_2$Se$_3$. The blue arrow marks an FeSe band feature at $E_{bin} \approx 0.2$ eV. **(b)** Fermi surface showing the two topological surface states. Both states display the expected hexagonal warping of the constant energy contour. The inset hexagon shows the orientation of the surface BZ of Bi$_2$Se$_3$. The dashed arrow indicates the direction of the cut in (a). **(c)** Constant-energy contour acquired at $h\nu = 53.6$ eV across a wider range of momenta, by integrating the spectral weight from a binding energy of +0.055 to -0.055 eV, revealing the orientation of the Bi$_2$Se$_3$ BZ (black dashed hexagons) and the BZs of the three domains of FeSe (green, purple and yellow dashed squares). The small black hexagons indicate the $\bar{\Gamma}$ and $\bar{\Gamma}'_{Bi2Se3}$ points and the small circles show the positions of the $\bar{M}_{FeSe}$ points.

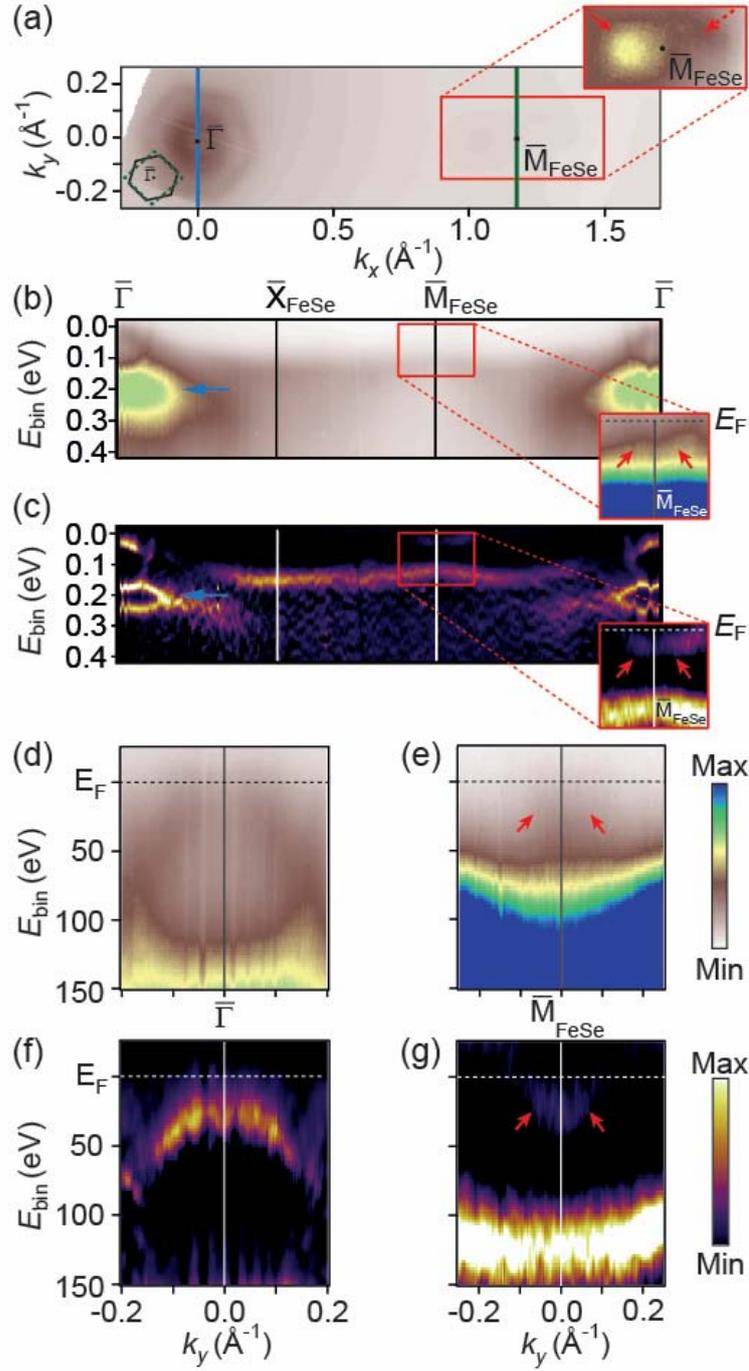

**Fig 3** (a) Photoemission intensity at the Fermi energy for FeSe/Bi$_2$Se$_3$, measured with photon energy $hv = 26$ eV. The $\bar{M}_{FeSe}$ point for one of the FeSe domains is visible in the measurement range, and is labeled. The red box encloses an extremely faint dumbbell-like feature at $\bar{M}_{FeSe}$; the inset shows this faint feature with higher contrast. (b) Photoemission intensity acquired at $hv = 53.6$ eV along the $\bar{\Gamma}$-$\bar{X}_{FeSe}$-$\bar{M}_{FeSe}$-$\bar{\Gamma}$ directions. (c) Curvature plot of (b). The red boxes in (b) and (c) enclose the extremely faint feature at $\overline{\text{M}}_{\text{FeSe}}$, and the insets show this feature magnified and at higher contrast. Blue arrows mark the location of the feature that is marked with an identical blue arrow in Fig. 2(a). (d), (e) Photoemission intensity along the cuts through $\overline{\Gamma}$ and $\overline{\text{M}}_{\text{FeSe}}$ that are marked with blue and green lines, respectively, in (a). (f), (g) Curvature plots obtained from (d) and (e), respectively. Red/blue arrows highlight the location of the feature at $\overline{\text{M}}_{\text{FeSe}}$. The orange arrow in Fig. 3(a) indicates the lobe of the dumbbell that falls in the second BZ, where the photoemission intensity is faint.

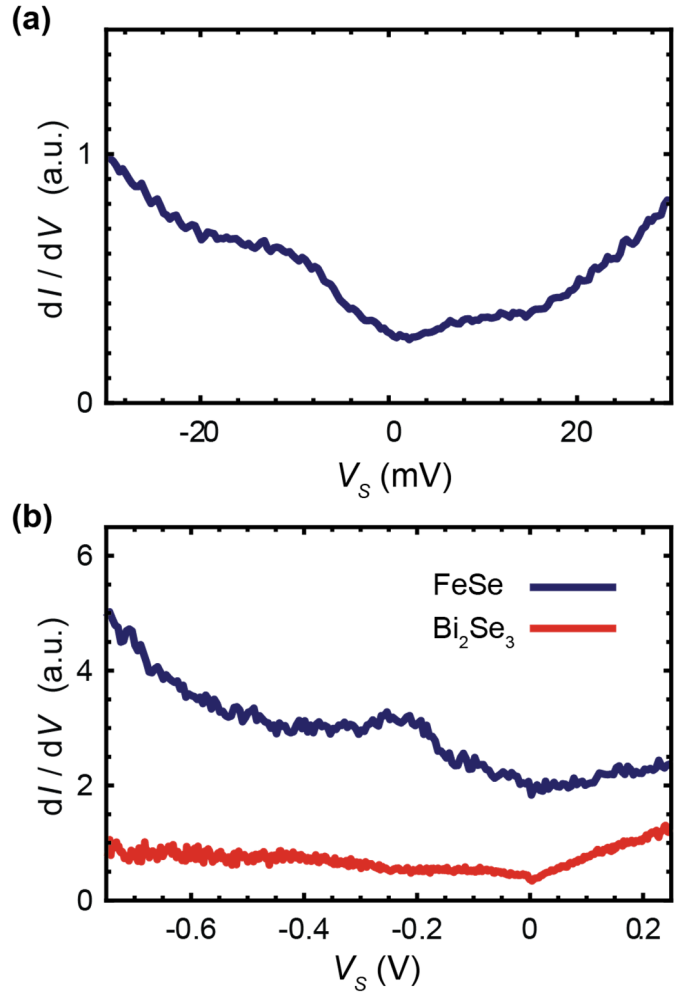

**Fig 4 (a)** A characteristic high resolution spectrum around the Fermi energy on an FeSe island with a minimum distance of 35 nm to the island edge. No superconducting gap is visible. $V_S = 50$ mV, $I_t = 500$ pA, $V_{mod} = 0.5$ mV. **(b)** Large range STS spectra taken on the $Bi_2Se_3$ (red) and the FeSe. The FeSe spectrum shows a peak at -225 meV, the same energy as that of the broad band in Fig 2. The blue spectrum is shifted for better visibility. Parameters, red: $V_S = 250$ mV, $I_t = 500$ pA, $V_{mod} = 0.5$ mV; blue: $V_S = 1$ V, $I_t = 500$ pA, $V_{mod} = 4$ mV.

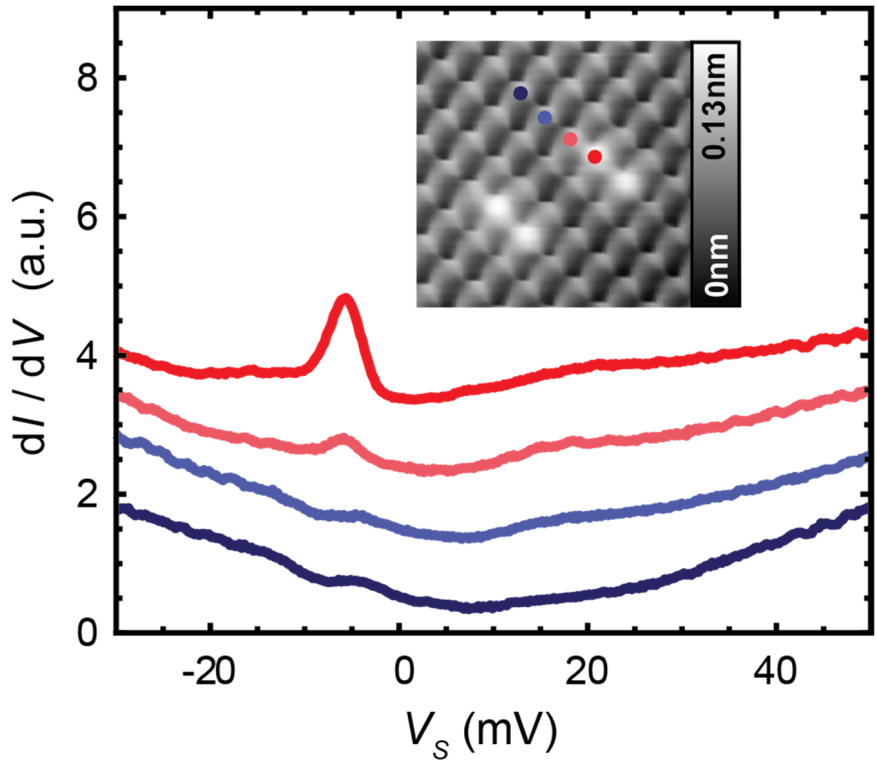

**Fig 5** Spectra taken on and close to a type A defect for an FeSe island. The inset shows the locations of the spectra in color code, from red (on the defect) to dark blue (three atomic lengths off the defect). The defect shows a resonance at -6 mV. $V_S$ = 50 mV, $I_t$ = 200 pA, $V_{mod}$ = 0.5 mV.